\documentclass[]{interact}

\usepackage{epstopdf}% To incorporate .eps illustrations using PDFLaTeX, etc.
\usepackage{subfigure}% Support for small, `sub' figures and tables

\usepackage[numbers,sort&compress,merge]{natbib}% Citation support using natbib.sty
\bibpunct[, ]{(}{)}{,}{n}{,}{,}% Citation support using natbib.sty
\theoremstyle{plain}% Theorem-like structures

\theoremstyle{definition}

\theoremstyle{remark}

\begin{document}

\title{A buffer gas beam source for short, intense and slow molecular pulses}

\author{
\name{S. Truppe, M. Hambach, S.M. Skoff, N.E. Bulleid,  J.S. Bumby,  R.J. Hendricks, E.A. Hinds, B. E. Sauer and M. R. Tarbutt\textsuperscript{a}\thanks{Email: s.truppe09@imperial.ac.uk}}
\affil{\textsuperscript{a}Centre for Cold Matter, Blackett Laboratory, Imperial College London, London SW7 2BW, United Kingdom}
}

\maketitle

\begin{abstract}
Experiments with cold molecules usually begin with a molecular source. We describe the construction and characteristics of a cryogenic buffer gas source of CaF molecules. The source emits pulses with a typical duration of 240~$\mu$s, a mean speed of about 150~m/s, and a flux of $5\times 10^{10}$ molecules per steradian per pulse in a single rotational state. 
\end{abstract}

\begin{keywords}
Buffer gas; molecular beam; radicals; laser cooling molecules; 
\end{keywords}

\section{Introduction}
A variety of applications in physics and chemistry call for molecules cooled to low temperatures~\cite{Bell2009, Stwalley2009, Carr2009, VandeMeerakker2012, Jankunas2015, Wall2016}. The starting point for many experiments with cold molecules is the pulsed molecular beam. For example, beams of cold molecules are used to test fundamental physics~\cite{Hudson2011, Baron2014, Shelkovnikov2008, Hudson2006a, Truppe2013a, Cheng2016, Cahn2014, Tokunaga2013} and to study collisions and reactions~\cite{Vogels2015, Jachymski2016, Klein2017}.

There are two main methods for producing pulsed beams of cold molecules, supersonic expansion and buffer gas cooling. In the first, the molecules of interest are entrained in a carrier gas which cools to low temperature via an adiabatic expansion from high to low pressure. The initial thermal energy of the carrier gas is converted to forward kinetic energy, resulting in fast-moving molecular beams. These beams have been used for a vast array of spectroscopic and collisional studies, and have been slowed using time-varying electric \cite{Bethlem1999} or magnetic fields~\cite{Narevicius2008} and subsequently trapped~\cite{Bethlem2000}. In the second method, the molecules are produced inside a cryogenically-cooled cell where they thermalise with a cold buffer gas of helium or neon and then exit through a hole in the cell to make a beam~\cite{Maxwell2005, Patterson2007, Patterson2009, Hutzler2011}. This method can produce slower beams that contain more molecules, though the number density may be lower because the pulse duration is typically much longer. Molecules in the low velocity tail of such a buffer gas beam can be selected \cite{Junglen2004, Sommer2010}, trapped \cite{Englert2011, Lu2014} and cooled \cite{Zeppenfeld2012}. Recently, buffer-gas-cooled beams of a few molecular species have been slowed to low velocity using radiation pressure~\cite{Barry2012, Zhelyazkova2014a, Yeo2015, Hemmerling2016, Truppe2017}, and these slow molecules have been captured and cooled to low temperatures in magneto-optical traps~\cite{Barry2014, Truppe2017a, Anderegg2017}. The high intensity and low velocity of the beams produced by the buffer gas method was crucial to the success of these experiments. Electric and magnetic deceleration techniques would benefit from these advantages too. However, these techniques usually accept only a small slice of the molecular pulse, preserving its density, and so the long pulses produced by a buffer gas source are unwelcome. In this paper, we describe a buffer gas source of CaF molecules that produces pulses with short duration, high intensity and low velocity. Our aim is to provide information that is useful to others who wish to build a similar source. We present the design of the source, give practical details about how to build it, and then characterise its performance.

\section{Design}

\begin{figure}[t]
	\centering
	\includegraphics{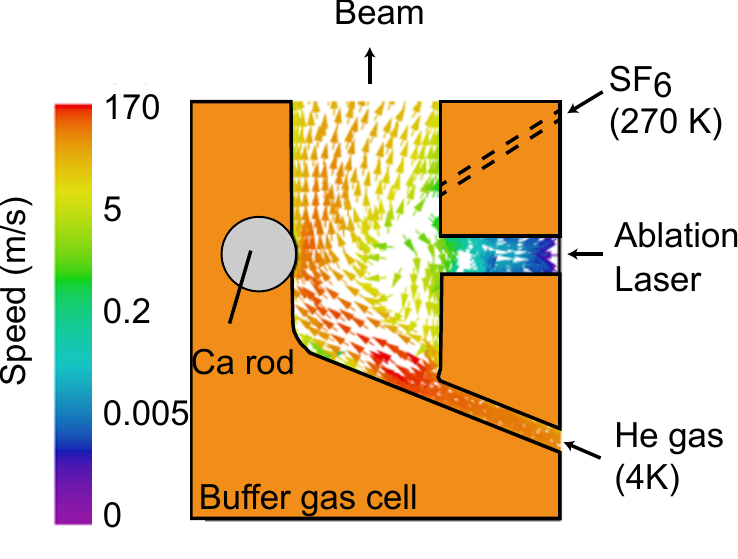}
	\caption{Design of buffer gas source showing the simulated flow of helium gas through the cell.~\cite{Bulleid2013}}
	\hspace{5pt}
	\label{cellSimulation}
\end{figure}

The working principles of buffer-gas-cooled beams have been reviewed extensively before~\cite{Hutzler2012}. We aimed to design a buffer-gas cell that would ensure efficient and rapid extraction of molecules produced inside by laser ablation. To achieve that, an objective was to minimise regions of low helium flow, since molecules entering such regions are likely to be lost, and to prevent vortex formation where molecules can be trapped~\cite{Bulleid2013}. Figure \ref{cellSimulation} illustrates the basic design of our source, and shows the flow of helium through the cell predicted using finite-element modelling software\footnote{Star-CCM+.}~\cite{Bulleid2013a}. The source has some features in common with those developed by Smalley and co-workers to produce supersonic metal cluster beams~\cite{Powers1982}, and others in common with buffer gas sources, so can be thought of as a hybrid of these two source types. We produce CaF inside the cell by laser ablation of a calcium rod in the presence of SF$_6$. Cold helium enters the cell through an angled tube, flows along the angled base of the cell and is directed towards the ablation target. This creates a region of high flow and high density near the target. The high density confines the ablation products, preventing them from reaching the cell walls, and also ensures rapid thermalization with the cold helium. The high flow results in rapid extraction of molecules from the cell, yielding a short pulse and leaving little time for diffusion to the walls. The simulation predicts that molecules will exit the cell with a mean velocity of 120~m/s in a pulse that has a full width at half maximum of 340~$\mu$s. We will see in section \ref{Sec:Characteristics} that the measured parameters are not too different from these.

\section{Construction details}

\begin{figure}[t]
 	\centering
 	\includegraphics[width=\textwidth]{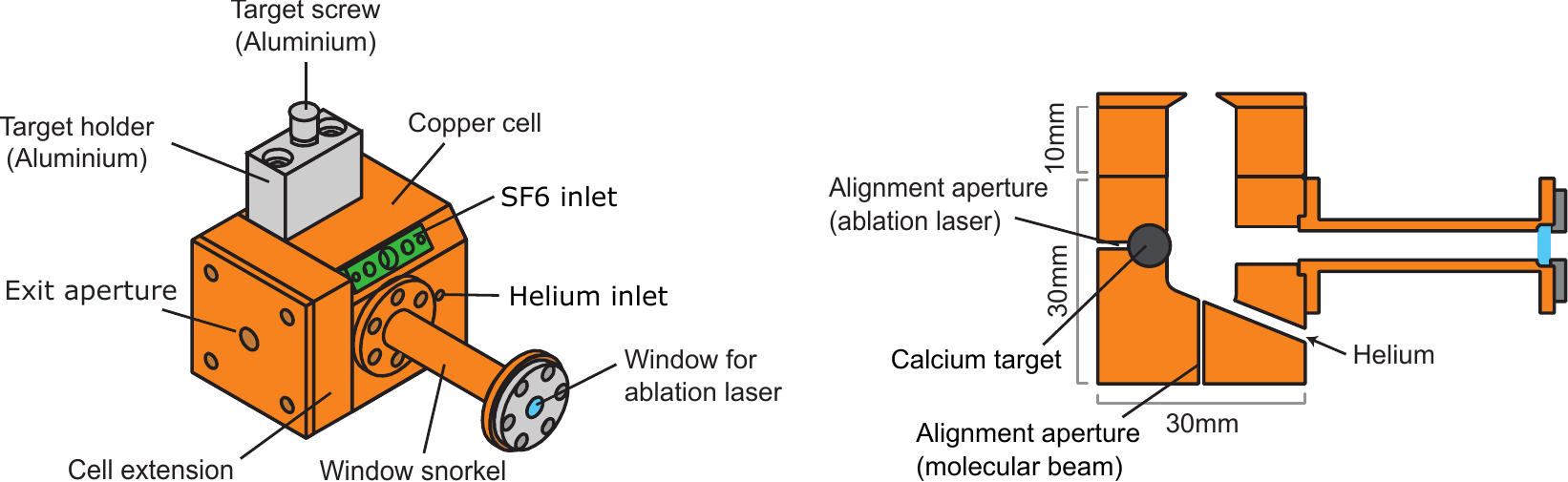}
 	\caption{Design details of the buffer gas cell.}
 	\hspace{5pt}
 	\label{cellDetail}
 \end{figure}

After experimenting with a few different versions of this basic design, we settled on the one shown in Figure~\ref{cellDetail}. This is the design we used for our recent experiments on laser slowing and magneto-optical trapping of CaF molecules~\cite{Truppe2017, Truppe2017a}. The length of the cell is extended by 10~mm, and we have added a conical aperture with an inner diameter of 3.5\,mm at the exit. The cell is discharge machined from an oxygen-free-copper block. It has a 10\,mm diameter bore and a base angled at $22^{\circ}$ to the horizontal where the cold helium gas enters. Warm SF$_6$ gas enters from a capillary attached to one edge of the cell. The ablation target is a Ca rod with a diameter of $5.9$\,mm and a length of $30$\,mm. We roughly cut the target from a Ca ingot and then turn it into a cylinder on a lathe. Once formed, the rod is cleaned with solvent and then stored under vacuum or mineral oil until needed. Just before the main vacuum chamber (see below) is closed, the target is polished with fine grit sand paper, attached to a finely-threaded aluminium screw, and inserted into the cell through a 6~mm diameter bore. A thin layer of Apiezon N vacuum grease is applied to the thread to prevent seizing. Rotating the screw rotates and translates the target, increasing the surface area that can be ablated and thus the lifetime of the target. For ablation, we use a Nd:YAG laser producing pulses of 5~ns duration and 5\,mJ energy at a wavelength of 1064~nm. We focus the output of the laser onto the target using a 75~cm focal length lens positioned about 55~cm from the target. The beam enters through a window offset from the cell by a narrow tube, which we call the snorkel. This helps to prevent ablation products coating the window.

\begin{figure}[t]
\centering
\includegraphics{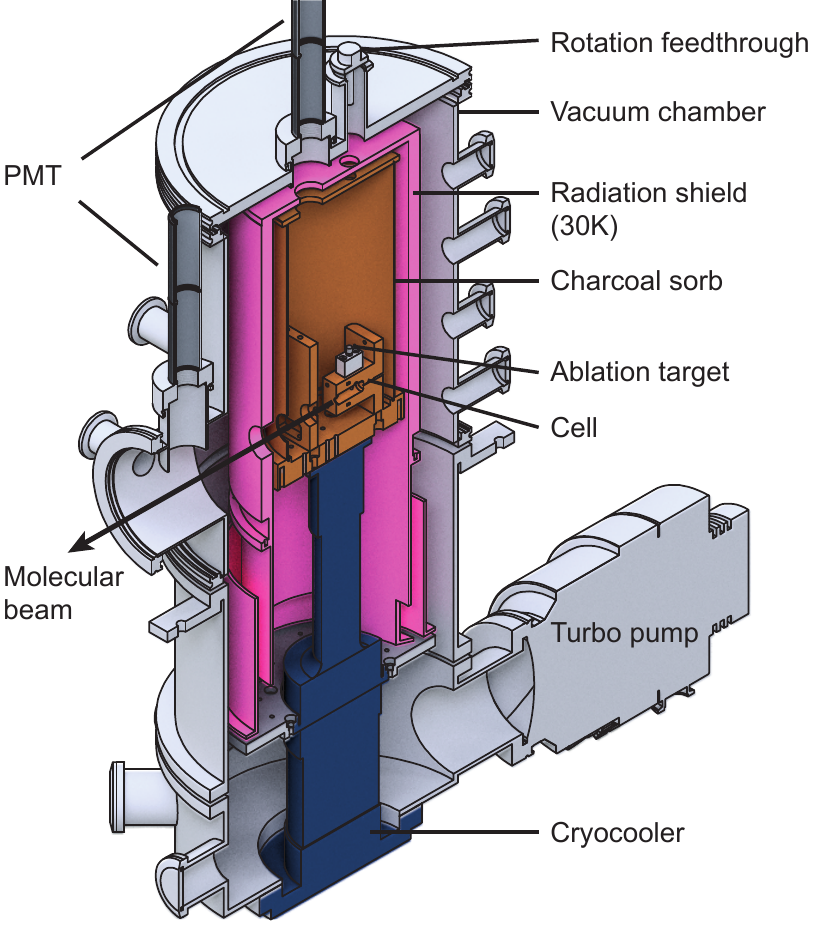}
\caption{Cut-away section through the centre of the source. The cryocooler head (blue) sits inside a vacuum chamber (grey). An aluminium radiation shield (pink) is attached to the first stage of the cryocooler. The copper inner cylinder (orange), attached to the second stage of the cryocooler, is coated with coconut charcoal and acts a sorb for helium. The buffer gas cell is attached to the second stage.}
\hspace{5pt}
\label{source3D}
\end{figure}

Figure \ref{source3D} shows a cut-away section through the centre of the source chamber. The cell is attached to a copper plate which is then mounted onto the cold head of a two stage Gifford-McMahon cryocooler\footnote{Sumitomo RDK-415D.}. The first and second stages of this cryocooler have cooling powers of 40~W at 50~K and 1.5~W at 4.2~K, respectively. Under our normal operating conditions, the second stage has a temperature of about 4~K. A vacuum chamber houses the cryocooler and provides ports for optical access and for the required gas, rotary and electrical feedthroughs. A turbomolecular pump with a pumping speed of 600~l/s for helium evacuates the chamber, reaching $10^{-7}$~mbar when the cryocooler is off.

An aluminium cylinder attached to the first stage of the cryocooler is cooled to about 30~K and serves as a radiation shield. This reduces the radiative heat load onto the second stage from approximately 50\,mW/cm$^2$ to below 0.05\,mW/cm$^2$. Aluminised mylar applied to the radiation shield can reduce the heat load further. Holes in the radiation shield let the ablation laser beam and gas pipes enter, and the molecular beam exit. Inside the radiation shield, surrounding the cell and attached to the 4~K cold head, is a copper cylinder. Coconut charcoal is glued to the inside of this cylinder using thermally conductive epoxy\footnote{Stycast 2850 FT with 23 LV catalyst.}. When cooled below 8~K, the charcoal acts as an efficient cryopump for helium gas \cite{Tobin1987, Sedgley1987}. The preparation and condition of the charcoal is important. We degrease the copper and paint on a thin layer of epoxy with a brush. The layer should be thin enough that most of the charcoal surface is exposed, but thick enough to provide good thermal conduction. We sprinkle on the charcoal, leave it to dry for a few hours, and finally bake the cylinder in air at 400~K for 24 hours. Once installed and under vacuum, we heat the cryopump to $\approx 330$~K for 48 hours. We have observed an improved performance of the pump after a few cycles of cooling down to 4~K and warming back up to 330~K.

Room temperature helium gas, with 99.999~\% purity, is fed into the vacuum chamber through 1/4-inch stainless-steel tubing. Inside the vacuum chamber, the stainless-steel tube is connected to a 1/4-inch copper tube which winds around a bobbin that is thermally anchored to the first stage of the cryocooler, cooling the gas to about 35~K. The tubing is soft-soldered to the bobbin along the length of the winding for good thermal contact. A second copper bobbin is attached to the second stage of the cryocooler, and copper tubing is wound and soft-soldered around this bobbin to cool the gas to about 4~K before it enters the cell. The end of this copper tube is soldered into the cell. The two bobbins are separated by a 20~cm long section of thin-walled 1/4-inch stainless steel tubing to minimise the thermal conductivity between them. To make each connection between copper and stainless-steel, we braze the copper tube into a stainless-steel Swagelok union and connect the stainless-steel tube to the union using a standard Swagelok connector. The brazed joints and Swagelok connectors are helium leak checked before closing the vacuum chamber. The helium flow is regulated with a flow controller\footnote{Alicat Scientific.}. When the flow rate is 1~sccm, the mean helium density in the cell is about $10^{15}$~cm$^{-3}$.

A novel feature of our cryogenic source is the formation of molecules by ablation of a pure metal in the presence of a reactant gas, SF$_{6}$ in this case~\cite{Bulleid2013a, Bumby2016}. We have found this to be better than using an ablation target formed from a mixture of powders. Lower energy ablation pulses can be used, the signal is more stable, and the target does not produce much of the dust that has been reported by others, which is thought to reduce the lifetime of molecules in buffer gas cells~\cite{Maxwell2007}. The room temperature SF$_6$ gas is fed into the vacuum chamber through a 1/4-inch stainless steel tube which connects to a copper capillary inside the vacuum chamber. This capillary is connected to the cell at the position shown in figure \ref{cellDetail}. At atmospheric pressure, the melting point of SF$_6$ is 222\,K. To prevent the gas from freezing, the capillary is thermally insulated from the cell by a polyimide (Vespel) spacer. The end of the capillary should be flush with the internal edge of the cell wall. If it is not inserted far enough SF$_6$ can freeze on the walls of the insulator or cell and clog the capillary. If it is inserted too far into the cell the hot tube heats the helium resulting in a hotter and faster molecular beam. A thermistor measures the temperature of the capillary close to the point where it enters the cell, and a heater wrapped around the capillary can be used to warm it up if the SF$_6$ freezes. In practice, the thermal isolation is good enough that the heater is never used.

At a distance of 30~mm from the exit aperture of the cell, the molecular beam passes through a 6~mm aperture in a charcoal-covered copper plate attached to the 4~K cold head. This reduces the helium gas load into the rest of the system, and is similar to the design in reference \cite{Barry2011}. Installing it reduced the pressure measured outside the radiation shields by a factor of 2.

We operate the source at a repetition rate of 2~Hz. The molecule flux gradually decays, so we rotate the target to expose fresh calcium after about $10^{4}$ shots. The variation in molecule number for various ablation spots is up to 50\%, but a spot with similar ablation yield can be found relatively quickly. The target is turned by the aluminium screw (see figure~\ref{cellDetail}) which is attached to a rotary vacuum feedthrough by a universal joint. The universal joint is thermally connected to the radiation shield via a bronze brush bushing and is thermally disconnected from the room temperature feedthrough and the cold aluminium screw using polyether ether ketone (PEEK) spacers. 

After about 24 hours of continuous operation, the cryopumps start to saturate. We regenerate them by warming them to a temperature exceeding 20~K. The helium then desorbs and is pumped away by the mechanical pumps. We also observe a significant reduction in the flux of molecules after about 20 hours of operation which cannot be recovered either by turning the target or regenerating the charcoal pumps. It may be that in this time enough SF$_6$ freezes to the cell walls to obstruct either the SF$_{6}$ input or the exit aperture of the cell. We find that we can recover the original flux by going through a heat up and cool down cycle overnight. We heat the source to at least 240~K in 6-7 hours using 50~W power resistors attached to the first and second stages of the cryocooler. We then immediately start cooling the source back down to 4~K. This takes approximately 6 hours so that the source is ready to use again the next morning. 

Early in the development of the source we noticed a gradual increase of the velocity of the molecular beam over weeks of operation. The thermal cycling described above did not fix this. Only by opening the source and cleaning the cell could we recover the original low beam velocity. We later discovered a small leak in one of the gas lines which was allowing air to leak into the cell. We hypothesise that this was oxidising the ablation products deposited on the cell walls. These oxidised products then form a thermally insulating layer which prevents the helium from fully thermalising (or re-thermalising \cite{Skoff2011}) with the cell walls, leading to a hotter and faster molecular beam.  Since fixing this small leak in the gas line we have been able to operate the source for many months without it requiring any maintenance.

\section{Characteristics}
\label{Sec:Characteristics}

The molecules are detected by cw laser-induced fluorescence detection on the $A^2\Pi_{1/2}(v=0, J=1/2)\leftarrow X^2\Sigma^+(v=0, N=1)$ transition near 494431\,GHz. This transition has four resolved components due to hyperfine and spin-rotation interactions. We apply radio-frequency sidebands to the probe laser in order to address all four components simultaneously~\cite{Zhelyazkova2014a, Truppe2017}. The probe laser beam is circular, has a total power of 0.5~mW and an intensity distribution with an rms radius of 1.5~mm. This probe beam intersects the molecular beam at $90^\circ$.

We first searched for a signal within the cryogenic region of the source, between 1.5 and 4~cm from the cell aperture. Laser-induced fluorescence from this region was imaged onto a photomultiplier tube (PMT) through a 1 inch window in the lid of the vacuum chamber and corresponding holes in the radiation shield and cryo-sorb (see figure~\ref{source3D}). After optimizing the signal at this position we probed the molecular beam further downstream, first at $53$\,cm and then at $130$\,cm. By comparing the signal inside and outside the cryogenic region, we can determine whether the molecular beam is attenuated by collisions with the helium buffer gas. This can be done either by calibrating the two detectors and then comparing the flux at the two positions, or by recording the flux as a function of the He flow rate inside and outside the cryogenic region. If the maximum molecular flux occurs at a lower flow rate at the downstream position than the upstream position, it indicates that the pumping speed for helium is insufficient. The source then has to be operated at a flow rate which does not extract the most molecules from the cell. In that case, more charcoal, and smaller apertures in the radiation shield and charcoal-covered plate are likely to improve the signal and stability of the source. 

\begin{figure}[t]
	\centering
	\includegraphics[width=\textwidth]{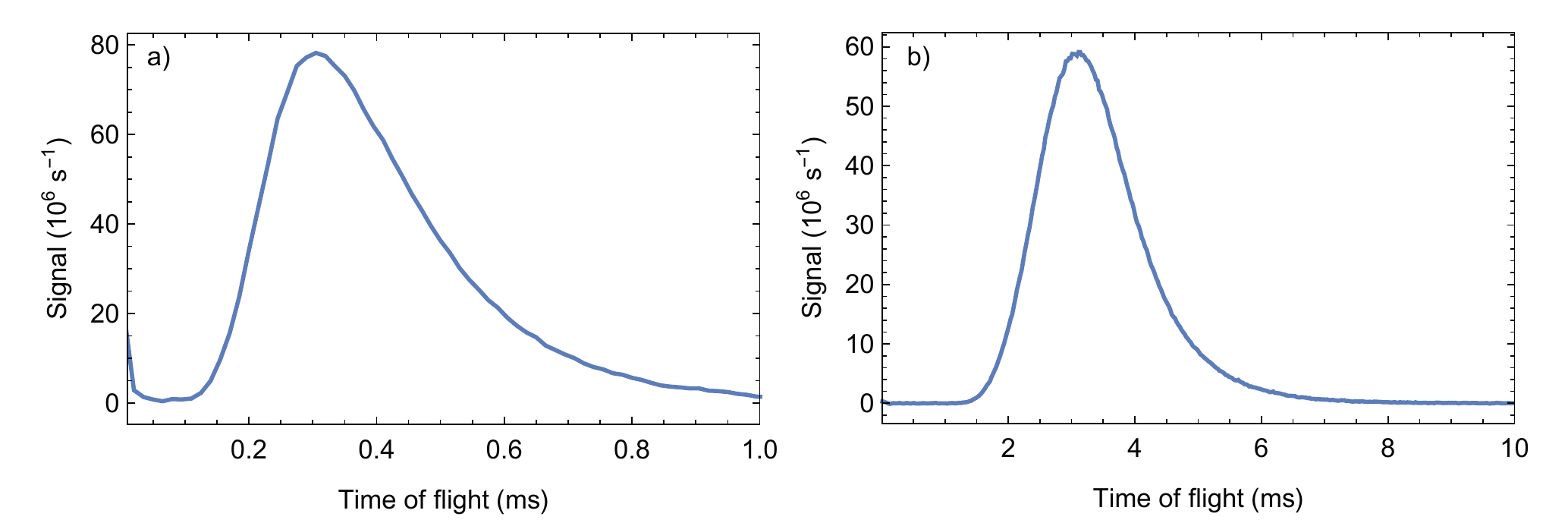}
	\caption{Typical time of flight profiles recorded at (a) 2.5~cm and (b) 53~cm from the cell aperture.}
	\hspace{5pt}
	\label{tofs}
\end{figure}

Figure \ref{tofs}(a) shows a typical time of flight profile recorded 2.5\,cm from the cell aperture. The molecular flux peaks 300~$\mu$s after the ablation laser fires. Accounting for the time taken to travel the 2.5~cm from the exit, we conclude that the molecules exit the cell about 100~$\mu$s after they are produced, showing that they are extracted rapidly. Modelling the interior volume of our cell as a cube of side length 1~cm, and using a typical diffusion cross section of $10^{-14}$~cm$^{2}$ and a typical helium density of $10^{15}$~cm$^{-3}$, we find that the diffusion time to the walls is about 100~$\mu$s. So we see that the extraction time and the diffusion time are similar. The pulse of molecules measured 2.5~cm from the aperture has a duration (full width at half maximum) of 275~$\mu$s. This is far shorter than other buffer-gas sources which typically emit pulses with durations from a few ms up to $\sim 100$~ms~\cite{Barry2011}. The short duration is a consequence of the small cell volume and rapid extraction of the molecules. Figure \ref{tofs}(b) shows a typical time of flight profile recorded 53~cm from the cell aperture. Here, the detection efficiency is much higher, which is why the signal rate is similar to that measured at the upstream detector. The signal peaks at 3.14~ms, indicating that the mean speed is about 170~m/s. At 4~K, an effusive beam of CaF would have a mean speed of 45~m/s, while a supersonic beam of helium has a speed of 204~m/s. In common with other buffer gas sources, the mean speeds we measure lie between these two extremes. It is possible to reach lower speeds by operating in a more effusive regime~\cite{Maxwell2005} or by using a two-stage buffer gas source~\cite{Lu2011}, but the flux obtained is considerably lower.

\begin{figure}[t]
	\centering
	\includegraphics{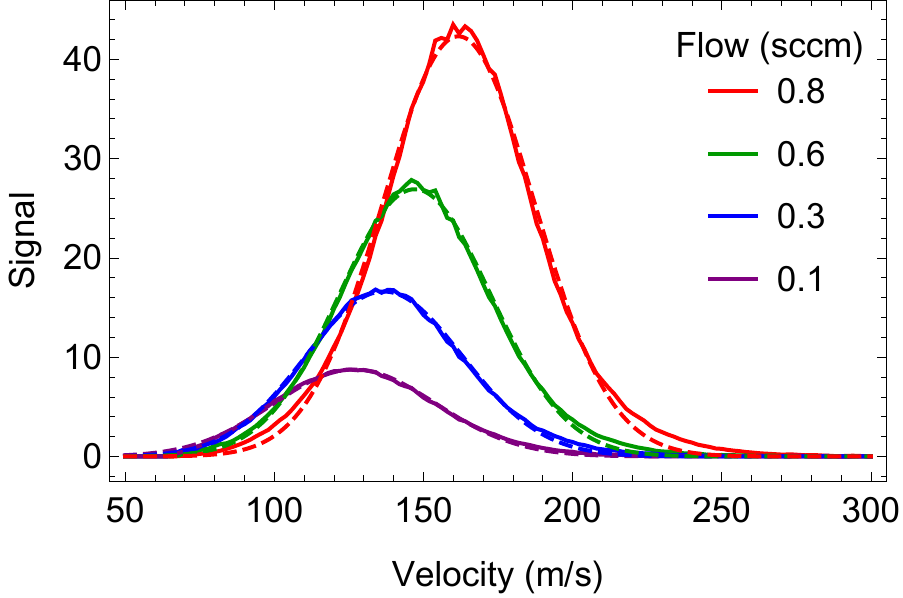}
	\caption{Solid lines: velocity distributions measured 130~cm from the source for four different helium flows. Dashed lines: fits to the model $f(v) = A v^{2}\exp[-M (v-v_{0})^{2}/(2k_{\rm B}T)]$.}
	\label{flows}
\end{figure}

Figure \ref{flows} shows velocity distributions for four different helium flows. These distributions are determined from the time-of-flight profiles measured a distance $L=130$~cm from the source, assuming the relation $v = L/t$ between the velocity $v$ and arrival time $t$. This relation is accurate when the arrival time is far greater than the duration of the pulse exiting the source, which is the case for these measurements. The method has been verified previously~\cite{Truppe2017}.  We see from figure~\ref{flows} that the signal increases with helium flow rate in the range 0.1-0.8~sccm. In this range, the number of molecules scales approximately linearly with the flow rate. At flows exceeding 1~sccm this number saturates. We expect this saturation to occur once the extraction time from the cell is shorter than the diffusion time to the walls, because then most of the molecules produced in the cell are extracted into the beam~\cite{Hutzler2011}. This occurs once the flow is greater than $\approx 1$~sccm.  The dashed lines in figure~\ref{flows} are fits to the model $f(v) = A v^{2}\exp[-M (v-v_{0})^{2}/(2k_{\rm B}T)]$, where $M$ is the mass of a CaF molecule, $A$ is an amplitude, $T$ is the temperature and $v_{0}$ is a velocity boost due to entrainment with the helium. The measured distributions tend to have more molecules at high velocity than this model, but otherwise the model describes the data well. The best-fit value of $v_{0}$ increases with flow, from 114~m/s at 0.1~sccm to 153~m/s at 0.8~sccm. The best-fit value of $T$ decreases from 5.8~K at 0.1~sccm to 4.8~K at 0.8~sccm. We see that higher flows produce higher velocities and lower temperatures, as a result of the flow regime shifting towards the supersonic limit and the molecules becoming more fully thermalised. Thus, while higher flows give more molecules in total, lower flows yield more molecules at the lowest velocities. 

We note that the translational temperature and the shape of the velocity distribution depend on the ablation energy and alignment of the ablation laser. We sometimes see distributions characterised by $T\approx 10$~K, having long tails extending to high velocities, indicating incomplete thermalization. We can understand this with a simple model. Taking a typical mean density in the cell of $10^{15}$~cm$^{-3}$ (for flows around 0.5~sccm) and a typical collision cross section of $10^{-14}$~cm$^{2}$, the mean free path is about 0.7~mm. Therefore, molecules have roughly 40 collisions before escaping from the cell, perhaps somewhat more since the density is higher in the vicinity of the target. After $n$ hard-sphere collisions with stationary buffer gas atoms of mass $m$, a molecule of initial energy $E_0$ has an energy $E_{n} = E_{0}(1-f)^{n}$ where $f=2M m/(m+M)^{2} = 0.12$. This gives $E_{40} = 0.006 E_{0}$. Given that the temperatures involved in laser ablation are often a few thousand kelvin~\cite{Davis1985}, this number of collisions is barely sufficient to cool the molecules to the helium temperature. For other species, especially heavier ones or ones that require higher ablation energies for their production, more collisions may be needed to cool the molecules, requiring a longer cell or a smaller exit aperture. For those constructing a similar source for a different molecule, we recommend experimenting with the dimensions of the cell.

We turn now to the determination of the flux of molecules. Let $N_{\rm mol}$ be the number of molecules in the X$^{2}\Sigma^{+}(v=0,N=1)$ state per unit solid angle per pulse, and $N_{\rm ph}$ be the number of photons detected per pulse at a detector a distance $L$ from the source. They are related by
\begin{equation}
N_{\rm ph} = \frac{N_{\rm mol}}{L^{2}} \iint \epsilon(x,y) p(y)\,dx\,dy.\nonumber
\end{equation}
Here, $x$ is the displacement from the centre of the detection system in the direction of the probe beam, $y$ is the displacement along the line to the PMT, $\epsilon(x,y)$ is the efficiency of detecting photons emitted at position $(x,y)$, and $p(y)$ is the number of photons emitted per molecule as it travels through the probe beam with displacement $y$. The fluorescence is collimated by an aspheric lens of diameter 50~mm and focal length 39~mm, and then imaged onto the PMT with a magnification of 2 by a second lens. A spherical mirror retro-reflects light emitted in the opposite direction to the PMT, doubling the collection efficiency. An aperture in the imaging plane ensures that only light from a small rectangular region can be detected. This region is 3~mm wide in the direction of the molecular beam and $w=5$~mm wide along $x$. The detection efficiency, $\epsilon$, is uniform across this small area, and the depth of field is large enough that $\epsilon$ is also independent of $y$ for all values of $y$ where photons are emitted. Taking into account the collection efficiency, PMT quantum efficiency and the transmission of lenses and windows, we estimate $\epsilon = 0.01$. With this simplification, we can write
\begin{equation}
N_{\rm mol} = \frac{N_{\rm ph}L^{2}}{\epsilon w \int p(y)\,dy}.
\label{Nmol}
\end{equation}  

At $L=53$~cm we detect $N_{\rm ph}=1.15\times 10^5$ photons per pulse. We use a rate equation model~\cite{Tarbutt2015} to calculate how many photons a molecule emits within the detection area as it travels through the probe beam, averaged over the distribution of velocities. Because the probe beam size is small, the integral rapidly converges as its limits are extended outwards. The calculation gives $\int p(y)\,dy = 19.6$~mm.  To verify this rate model calculation, we measure how $N_{\rm ph}$ changes with probe laser power, and how it changes between a single frequency probe and a probe with rf sidebands added (see above), then compare these results to the predictions of the model. They are in good qualitative agreement, but the comparisons indicate that, for the parameters used in our estimate of $N_{\rm mol}$, the model overestimates $\int p(y)\,dy$ by about 50\%. We think the scattering rate may have been limited by optical pumping into dark states which were not effectively de-stabilized by the small magnetic field in the probe region, an effect not captured by the rate model. Therefore, we take $\int p(y)\,dy = 13$~mm. Using these values in equation (\ref{Nmol}) we obtain $N_{\rm mol} = 5\times 10^{10}$ per steradian per shot, accurate to within a factor of 2. This flux is similar to that obtained from other buffer gas sources where molecules are produced by laser ablation~\cite{Barry2011, Hutzler2011a}. 

The source is very stable when operated at a repetition rate of 2~Hz. The standard deviation of the flux measured over $10^{4}$ shots is 12\% of the mean flux. This reduces to 6~\% when the shots are phase-locked to the compression cycle of the cryocooler. Increasing the repetition rate to 5~Hz reduces the mean number of molecules per shot by a factor of 2 and results in a very erratic flux with far higher shot-to-shot fluctuations. At 10~Hz the signal decays very rapidly, reaching about 2\% of its initial value after 500 shots. It is possible that, at higher repetition rates, local heating of the target produces a molten surface that is not amenable to ablation. We are currently investigating the cause with a view to running sources of this kind at higher repetition rates.

\section{Conclusions}

We have described a cryogenic buffer gas source that produces pulses of CaF molecules with a typical duration of 240~$\mu$s, a mean speed of about 150~m/s, and a flux of $5\times 10^{10}$ molecules per steradian per pulse in a single rotational state. The flux diminishes after running the source continuously for about 24 hours, but is easily recovered by going through a warm-up and cool-down cycle that takes about 12 hours. The source has run for many months without requiring maintenance. It has been an invaluable starting point for making cold, slow, velocity-controlled beams by frequency-chirped laser slowing~\cite{Truppe2017}, and for loading a magneto-optical trap of CaF~\cite{Truppe2017a}. As well as these applications, the source may be useful for spectroscopic studies and for loading Stark and Zeeman decelerators. The low velocity and high density of the pulses produced will help to extend the applicability of these deceleration devices to heavier molecules~\cite{Bulleid2012, Mathavan2016}. By providing slower, more intense beams, the source could be used to improve experiments that test fundamental physics, such as the measurement of the electric dipole moment of the electron or proton~\cite{Hudson2011, Hunter2012}, or studies of nuclear parity violation using molecules~\cite{Cahn2014}.

\section{Dedication}

We dedicate this paper to Danny Segal who was a good friend and brilliant colleague to many of us. Danny helped us to get started on buffer gas cooling and worked with us on early versions of our sources~\cite{Skoff2009, Skoff2011}. As often happens with new projects we had plenty of enthusiasm but no equipment, no funding, and little expertise. To get going, we needed a helium cryostat. Danny went for a walk around the department and returned an hour or two later with two perfectly functional cryostats. He was endlessly generous, and he brought out the generosity in others. Unlike the rest of us, Danny knew a bit about cryogenics, so he helped us modify one of the cryostats and taught us how to use it. We used it for all our early work on buffer gas cooling. Scientific progress was a source of joy for Danny. He would have been delighted to see how those first endeavours cobbled together from borrowed parts, requiring constant maintenance, tinkering and botching, led to the trouble-free source described in this paper and to the new science with ultracold molecules that has now become feasible. 

\section*{Acknowledgements}
We are grateful to John Dyne, Giovanni Marinaro and Valerijus Gerulis for technical assistance. 

\section*{Funding}
This work was supported by the EPSRC under grants EP/I012044, EP/M027716 and EP/P01058X/1, and by the European
Research Council under the European Union’s Seventh Framework
Programme (FP7/2007-2013)/ ERC grant agreement 320789.

%\section{References}
%\begin{thebibliography}{99}
%\bibliographystyle{tfp}
%\bibliography{bib}

\begin{thebibliography}{99}

\bibitem{Bell2009}
Bell, M.T.; {P. Softley}, T. {Ultracold molecules and ultracold chemistry},
  \emph{Molecular Physics}  \textbf{2009}, \emph{107}~(2), 99--132.


\bibitem{Stwalley2009}
Stwalley, W.C., Krems, R.V., Friedrich, B., Eds. \emph{{Cold Molecules}}; CRC
  Press, 2009.

\bibitem{Carr2009}
Carr, L.D.; DeMille, D.; Krems, R.V.; Ye, J. {Cold and ultracold molecules:
  Science, technology and applications}, \emph{New Journal of Physics}
  \textbf{2009}, \emph{11}, 055049.

\bibitem{VandeMeerakker2012}
van~de Meerakker, S.Y.T.; Bethlem, H.L.; Vanhaecke, N.; Meijer, G. {Manipulation
  and control of molecular beams}, \emph{Chemical Reviews}  \textbf{2012},
  \emph{112}~(9), 4828--4878.

\bibitem{Jankunas2015}
Jankunas, J.; Osterwalder, A. {Cold and Controlled Molecular Beams: Production
  and Applications}, \emph{Annual Review of Physical Chemistry}  \textbf{2015},
  \emph{66}~(1), 241--262.

\bibitem{Wall2016}
Wall, T.E. {Preparation of cold molecules for high-precision measurements},
  \emph{Journal of Physics B}
  \textbf{2016}, \emph{49}~(24), 243001.
	
\bibitem{Hudson2011}
Hudson, J.J.; Kara, D.M.; Smallman, I.J.; Sauer, B.E.; Tarbutt, M.R.; Hinds,
  E.A. {Improved measurement of the shape of the electron}, \emph{Nature}
  \textbf{2011}, \emph{473}~(7348), 493--496.

\bibitem{Baron2014}
Baron, J.; Campbell, W.C.; DeMille, D.; Doyle, J.M.; Gabrielse, G.; Gurevich,
  Y.V.; Hess, P.W.; Hutzler, N.R.; Kirilov, E.; Kozyryev, I.; et~al. {Order of
  Magnitude Smaller Limit on the Electric Dipole Moment of the Electron},
  \emph{Science}  \textbf{2014}, \emph{343}~(6168), 269--272.

\bibitem{Shelkovnikov2008}
Shelkovnikov, A.; Butcher, R.J.; Chardonnet, C.; Amy-Klein, A. {Stability of
  the proton-to-electron mass ratio}, \emph{Physical Review Letters}
  \textbf{2008}, \emph{100}~(15).

\bibitem{Hudson2006a}
Hudson, E.R.; Lewandowski, H.J.; Sawyer, B.C.; Ye, J. {Cold molecule
  spectroscopy for constraining the evolution of the fine structure constant},
  \emph{Physical Review Letters}  \textbf{2006}, \emph{96}~(14), 1--4.

\bibitem{Truppe2013a}
Truppe, S.; Hendricks, R.J.; Tokunaga, S.K.; Lewandowski, H.J.; Kozlov, M.G.; Henkel,
  C.; Hinds, E.A.; Tarbutt, M.R. {A search for varying fundamental constants using
  hertz-level frequency measurements of cold CH molecules}, \emph{Nature
  Communications}  \textbf{2013}, \emph{4}, 1--7.

\bibitem{Cheng2016}
Cheng, C.; {Van Der Poel}, A.P.P.; Jansen, P.; Quintero-P\'{e}rez, M.; Wall, T.E.;
  Ubachs, W.; Bethlem, H.L. {Molecular Fountain}, \emph{Physical Review
  Letters}  \textbf{2016}, \emph{117}~(25), 1--5.

\bibitem{Cahn2014}
Cahn, S.; Ammon, J.; Kirilov, E.; Gurevich, Y.; Murphree, D.; Paolino, R.;
  Rahmlow, D.; Kozlov, M.; DeMille, D. {Zeeman-Tuned Rotational Level-Crossing
  Spectroscopy in a Diatomic Free Radical}, \emph{Physical Review Letters}
  \textbf{2014}, \emph{112}~(16), 163002.

\bibitem{Tokunaga2013}
Tokunaga, S.K.; Stoeffler, C.; Auguste, F.; Shelkovnikov, A.; Daussy, C.;
  Amy-Klein, A.; Chardonnet, C.; Darqui\'{e}, B. {Probing weak force-induced parity
  violation by high-resolution mid-infrared molecular spectroscopy},
  \emph{Molecular Physics}  \textbf{2013}, \emph{111}~(14-15), 2363--2373.

\bibitem{Henson2012}
Henson, A.B.; Gersten, S.; Shagam, Y.; Narevicius, J.; Narevicius, E.
  {Observation of Resonances in Penning Ionization Reactions at Sub-Kelvin
  Temperatures in Merged Beams}, \emph{Science}  \textbf{2012},
  \emph{338}~(6104), 234--238.

\bibitem{Vogels2015}
Vogels, S.N.; Onvlee, J.; Chefdeville, S.; van~der Avoird, A.; Groenenboom,
  G.C.; van~de Meerakker, S.Y.T. {Imaging resonances in low-energy NO-He
  inelastic collisions}, \emph{Science}  \textbf{2015}, \emph{350}, 787--790.

\bibitem{Jachymski2016}
Jachymski, K.; Hapka, M.; Jankunas, J.; Osterwalder, A. {Experimental and
  theoretical studies of low-energy Penning ionization of NH$_{3}$, CH$_{3}$F
  and CHF$_{3}$}, \emph{Chem. Phys. Chem.}  \textbf{2016}, \emph{17}, 3776.

\bibitem{Klein2017}
Klein, A.; Shagam, Y.; Skomorowski, W.; \.Zuchowski, P.S.; Pawlak, M.; Janssen,
  L.M.C.; Moiseyev, N.; van~de Meerakker, S.Y.T.; van~der Avoird, A.; Koch,
  C.P.; et~al. {Directly probing anisotropy in atom-molecule collisions through
  quantum scattering resonances}, \emph{Nature Physics}  \textbf{2017},
  \emph{13}, 35--38.

\bibitem{Bethlem1999}
Bethlem, H.; Berden, G.; Meijer, G. {Decelerating Neutral Dipolar Molecules},
  \emph{Physical Review Letters}  \textbf{1999}, \emph{83}~(8), 1558--1561.

\bibitem{Narevicius2008}
Narevicius, E.; Libson, A.; Parthey, C.G.; Chavez, I.; Narevicius, J.; Even,
  U.; Raizen, M.G. {Stopping supersonic oxygen with a series of pulsed
  electromagnetic coils: A molecular coilgun}, \emph{Physical Review A}  \textbf{2008}, \emph{77}~(5), 1--4.

\bibitem{Bethlem2000}
Bethlem, H.L.; Berden, G.; Crompvoets, F.M.H.; Jongma, R.T.; van Roij, A.J.a.;
  Meijer, G. {Electrostatic trapping of ammonia molecules}, \emph{Nature}
  \textbf{2000}, \emph{406}~(August), 491--495.

\bibitem{Maxwell2005}
Maxwell, S.E.; Brahms, N.; Decarvalho, R.; Glenn, D.R.; Helton, J.S.; Nguyen,
  S.V.; Patterson, D.; Petricka, J.; DeMille, D.; Doyle, J.M. {High-flux beam
  source for cold, slow atoms or molecules}, \emph{Physical Review Letters}
  \textbf{2005}, \emph{95}~(17), 1--4.

\bibitem{Patterson2007}
Patterson, D.; Doyle, J.M. {Bright, guided molecular beam with hydrodynamic
  enhancement}, \emph{Journal of Chemical Physics}  \textbf{2007},
  \emph{126}~(15), 1--5.

\bibitem{Patterson2009}
Patterson, D.; Rasmussen, J.; Doyle, J.M. {Intense atomic and molecular beams
  via neon buffer-gas cooling}, \emph{New Journal of Physics}  \textbf{2009},
  \emph{11}.

\bibitem{Hutzler2011}
Hutzler, N.R.; Lu, H.I.; Doyle, J.M. {The Buffer Gas Beam: An Intense, Cold,
  and Slow Source for Atoms and Molecules}, \emph{Chemical Reviews}
  \textbf{2012}, \emph{112}~(9), 4803--4827.

\bibitem{Junglen2004}
Junglen, T.; Rieger, T.; Rangwala, S.A.; Pinkse, P.W.H.; Rempe, G. {Slow
  ammonia molecules in an electrostatic quadrupole guide}, \emph{European
  Physical Journal D}  \textbf{2004}, \emph{31}~(2), 365--373.

\bibitem{Sommer2010}
Sommer, C.; Motsch, M.; Chervenkov, S.; {Van Buuren}, L.D.; Zeppenfeld, M.;
  Pinkse, P.W.H.; Rempe, G. {Velocity-selected molecular pulses produced by an
  electric guide}, \emph{Physical Review A}  \textbf{2010}, \emph{82}~(1), 1--7.

\bibitem{Englert2011}
Englert, B.G.U.; Mielenz, M.; Sommer, C.; Bayerl, J.; Motsch, M.; Pinkse,
  P.W.H.; Rempe, G.; Zeppenfeld, M. {Storage and adiabatic cooling of polar
  molecules in a microstructured trap}, \emph{Physical Review Letters}
  \textbf{2011}, \emph{107}~(26), 1--4.

\bibitem{Lu2014}
Lu, H.I.; Kozyryev, I.; Hemmerling, B.; Piskorski, J.; Doyle, J.M. {Magnetic
  trapping of molecules via optical loading and magnetic slowing},
  \emph{Physical Review Letters}  \textbf{2014}, \emph{112}, 113006.

\bibitem{Zeppenfeld2012}
Zeppenfeld, M.; Englert, B.G.U.; Gl{\"{o}}ckner, R.; Prehn, A.; Mielenz, M.;
  Sommer, C.; van Buuren, L.D.; Motsch, M.; Rempe, G. {Sisyphus cooling of
  electrically trapped polyatomic molecules}, \emph{Nature}  \textbf{2012},
  \emph{491}~(7425), 570--573.

\bibitem{Barry2012}
Barry, J.F.; Shuman, E.S.; Norrgard, E.B.; Demille, D. {Laser radiation
  pressure slowing of a molecular beam}, \emph{Physical Review Letters}
  \textbf{2012}, \emph{108}~(10), 1--5.

\bibitem{Zhelyazkova2014a}
Zhelyazkova, V.; Cournol, A.; Wall, T.E.; Matsushima, A.; Hudson, J.J.; Hinds,
  E.A.; Tarbutt, M.R.; Sauer, B.E. {Laser cooling and slowing of CaF
  molecules}, \emph{Physical Review A}  \textbf{2014}, \emph{89}~(5), 053416.
	
\bibitem{Yeo2015}
Yeo, M.; Hummon, M.T.; Collopy, A.L.; Yan, B.; Hemmerling, B.; Chae, E.; Doyle,
  J.M.; Ye, J. {Rotational State Microwave Mixing for Laser Cooling of Complex
  Diatomic Molecules}, \emph{Physical Review Letters}  \textbf{2015},
  \emph{114}~(22), 1--5.

\bibitem{Hemmerling2016}
Hemmerling, B.; Chae, E.; Ravi, A.; Anderegg, L.; Drayna, G.K.; Hutzler, N.R.;
  Collopy, A.L.; Ye, J.; Ketterle, W.; Doyle, J.M. {Laser slowing of CaF
  molecules to near the capture velocity of a molecular MOT}, \emph{Journal of
  Physics B: Atomic, Molecular and Optical Physics}  \textbf{2016},
  \emph{49}~(17), 174001.

\bibitem{Truppe2017}
Truppe, S.; Williams, H.J.; Fitch, N.J.; Hambach, M.; Wall, T.E.; Hinds, E.A.;
  Sauer, B.E.; Tarbutt, M.R. {An intense, cold, velocity-controlled molecular
  beam by frequency-chirped laser slowing}, \emph{New Journal of Physics}
  \textbf{2017}, \emph{19}~(2), 022001.

\bibitem{Barry2014}
Barry, J.F.; McCarron, D.J.; Norrgard, E.B.; Steinecker, M.H.; DeMille, D.
  {Magneto-optical trapping of a diatomic molecule}, \emph{Nature}
  \textbf{2014}, \emph{512}~(7514), 286--289.

\bibitem{Truppe2017a}
Truppe, S.; Williams, H.J.; Hambach, M.; Caldwell, L.; Fitch, N.J.; Hinds,
  E.A.; Sauer, B.E.; Tarbutt, M.R. {Molecules cooled below the Doppler limit}, \emph{arXiv:1706.07848} \textbf{2017}.

\bibitem{Anderegg2017}
Anderegg, L.; Augenbraun, B.; Chae, E.; Hemmerling, B.; Hutzler, N.R.; Ravi,
  A.; Collopy, A.; Ye, J.; Ketterle, W.; Doyle, J. {Radio Frequency
  Magneto-Optical Trapping of CaF with High Density},  \emph{arXiv:1705.10288} \textbf{2017}.

\bibitem{Hutzler2012}
Hutzler, N.R.; Lu, H.I.; Doyle, J.M. {The buffer gas beam: An intense, cold,
  and slow source for atoms and molecules}, \emph{Chemical Reviews}
  \textbf{2012}, \emph{112}~(9), 4803--4827.

\bibitem{Bulleid2013}
Bulleid, N.E.; Skoff, S.M.; Hendricks, R.J.; Sauer, B.E.; Hinds, E.A.; Tarbutt,
  M.R. {Characterization of a cryogenic beam source for atoms and molecules},
  \emph{Physical Chemistry Chemical Physics}  \textbf{2013}, \emph{15}~(29),
  12299.
  
\bibitem{Bulleid2013a}
Bulleid, N.E. {Slow, cold beams of polar molecules for precision measurements},\emph{PhD thesis}, Imperial College London, 2013.

\bibitem{Powers1982}
Powers, D.E.; Hansen, S.G.; Geusic, M.E.; Pulu, A.C.; Hopkins, J.B.; Dietz, T.G.; Duncan, M.A.; Langridge-Smith, P.R.R.; Smalley, R.E.; {Supersonic metal cluster beams: laser photoionization studies of Cu$_2$}, \emph{Journal of Physical Chemistry} \textbf{1982}, \emph{86}, 2556-2560.

\bibitem{Tobin1987}
Tobin, A.G.; Batzer, D.W.S.H.; Call, W.R.; Tobin, A.G.; Sedgley, D.W.; Batzer,
  T.H.; Call, W.R. {Evaluation of charcoal sorbents for helium cryopumping in
  fusion reactors Evaluation of charcoal sorbents for helium cryopumping in
  fusion reactors}, \emph{Journal of Vacuum Science {\&} Technology A: Vacuum,
  Surfaces, and Films}  \textbf{1987}, \emph{5}~(1), 101.

\bibitem{Sedgley1987}
Sedgley, D.W.; Batzer, A.G.T.H.; Call, W.R.; Sedgley, D.W.; Tobin, A.G.
  {Characterization of charcoals for helium cryopumping in fusion devices
  Characterization of charcoals for helium cryopumping in fusion devices},
  \emph{Journal of Vacuum Science {\&} Technology A: Vacuum, Surfaces, and
  Films}  \textbf{1987}, \emph{2572}~(5).

\bibitem{Bumby2016}
Bumby, J. S. {Progress towards a source of cold, slow molecules
	for tests of fundamental physics}, \emph{PhD thesis}, Imperial College London, 2016.

\bibitem{Maxwell2007}
Maxwell, S.E. {Buffer gas cooled atoms and molecules: production, collisional studies and applications}, \emph{PhD thesis}, Harvard University, 2007.

\bibitem{Barry2011}
Barry, J.F.; Shuman, E.S.; DeMille, D. {A bright, slow cryogenic molecular beam
  source for free radicals}, \emph{Physical Chemistry Chemical Physics}
  \textbf{2011}, \emph{13}~(42), 18936.

\bibitem{Skoff2011}
Skoff, S.M.; Hendricks, R.J.; Sinclair, C.D.J.; Hudson, J.J.; Segal, D.M.;
  Sauer, B.E.; Hinds, E.A.; Tarbutt, M.R. {Diffusion, thermalization, and
  optical pumping of YbF molecules in a cold buffer-gas cell}, \emph{Physical
  Review A}  \textbf{2011}, \emph{83}~(2), 023418.


\bibitem{Lu2011}
Lu, H.I.; Rasmussen, J.; Wright, M.J.; Patterson, D.; Doyle, J.M. {A cold and
  slow molecular beam},  \emph{Physical Chemistry Chemical Physics} \textbf{2011}, \emph{13} 18986--18990.

\bibitem{Davis1985}
Davis, G.M.; Gower, M.C.; Fotakis, C.; Efthimiopoulos, T.; Argyrakis, P. {Spectroscopic studies of ArF laser photoablation of PMMA}, \emph{Applied Physics A} \textbf{1985}, \emph{36}, 27--30.

\bibitem{Tarbutt2015}
Tarbutt, M.R. {Magneto-optical trapping forces for atoms and molecules with
  complex level structures}, \emph{New Journal of Physics}  \textbf{2015}, \emph{17}~(1),
  15007.

\bibitem{Hutzler2011a}
Hutzler, N.R.; Parsons, M.F.; Gurevich, Y.V.; Hess, P.W.; Petrik, E.; Spaun,
  B.; Vutha, A.C.; DeMille, D.; Gabrielse, G.; Doyle, J.M. {A cryogenic beam of
  refractory, chemically reactive molecules with expansion cooling},
  \emph{Physical Chemistry Chemical Physics}  \textbf{2011}, \emph{13}~(42),
  18976.

\bibitem{Bulleid2012}
Bulleid, N.E.; Hendricks, R.J.; Hinds, E.A.; Meek, S.A.; Meijer, G.;
  Osterwalder, A.; Tarbutt, M.R. {Traveling-wave deceleration of heavy polar
  molecules in low-field-seeking states}, \emph{Physical Review A}
  \textbf{2012}, \emph{86}, 021404(R).

\bibitem{Mathavan2016}
Mathavan, S.C.; Zapara, A.; Esajas, Q.; Hoekstra, S. {Deceleration of a
  supersonic beam of SrF molecules to 120~m~s$^{-1}$}, \emph{ChemPhysChem}
   \textbf{2016}, \emph{17}, 3709--3713.

\bibitem{Hunter2012}
Hunter, L.R.; Peck, S.K.; Greenspon, A.S.; Alam, S.S.; DeMille, D. {Prospects
  for laser cooling TlF}, \emph{Physical Review A}  \textbf{2012}, \emph{85}~(1), 1--6.

\bibitem{Skoff2009}
Skoff, S.M.; Hendricks, R.J.; Sinclair, C.D.J.; Tarbutt, M.R.; Hudson, J.J.;
  Segal, D.M.; Sauer, B.E.; Hinds, E.A. {Doppler-free laser spectroscopy of
  buffer-gas-cooled molecular radicals}, \emph{New Journal of Physics}
  \textbf{2009}, \emph{11}, 123026.

\end{thebibliography}
%\end{thebibliography}

\end{document}